\documentclass[10pt,twocolumn]{article} 
\usepackage{times}
\usepackage{comment}
\usepackage{color}
\usepackage{overcite}
\usepackage{graphicx}
\usepackage{amsmath,amssymb,bm,braket}
\usepackage[top=20truemm,bottom=20truemm,left=15truemm,right=15truemm]{geometry}
\title{
{\Large{\bf
A New Method to Calculate 
a 2D Ising Universality Transition Point 
: Application near 
the Ashkin-Teller Multicritical Point
}}}
\author{Shunji Moriya$^1$\thanks{
	s.moriya@stat.phys.kyushu-u.ac.jp
	},
 and Kiyohide Nomura$^1$
\thanks{
	knomura@stat.phys.kyushu-u.ac.jp
	}\\
	{\it
$^1$Department of Physics, 
Kyushu University, 
Fukuoka 819-0395, Japan
}}
\date{}
\begin{document}
\maketitle
\begin{abstract}
		We propose a new method 
	to numerically calculate
	transition points that belongs
	to 2D Ising universality class 
	for quantum spin models.
	Generally, 
	near the multicritical point, 
	in conventional methods,
	a finite size correction 
	becomes very large.
 	To suppress the effect 
	of the multicritical point,
	we use 
	a z-axis twisted 
	boundary condition 
	and 
	a y-axis twisted 
	boundary condition. 
	We apply our method to 
	an $S=\frac{1}{2}$ 
	bond-alternating XXZ model. 
	The multicritical point of 
	this model 
	has a BKT transition, 
	where 
	the correlation length 
	diverges singularly.
	However, with our method, 
	the convergence of calculation 
	is highly improved, thus 
	we can calculate the transition point 
	even near the multicritical point.

\end{abstract}
	\section{Introduction}
	Critical phenomena are one of 
the important subjects in 
condensed matter physics. 
As a typical solvable model, 
a classical 2D Ising model is studied widely. 
\cite{onsager}
In the some limit, 
the transfer matrix of 
a classical 2D Ising model becomes  
a quantum 1D Transverse-Field Ising 
(TFI) model.
\cite{ising1}
\cite{ising2}
Several methods are proposed to calculate 
the 2D Ising universality transition points 
of quantum spin models.
\cite{phrg}\cite{dmrgbaxxz} 
But, 
when the model has a multicritical point, 
the scaling behaviors become difficult 
due to the effect of multiple critical lines.
So, conventional methods are not useful 
near a multicritical point.

Another method to calculate a transition point,
a {\it Level Spectroscopy} (LS) method  is 
useful to cancel logarithmic corrections of 
a Berezinskii-Kosterlitz-Thouless (BKT) transition. 
\cite{lsgau,lsbkt,lss1,lss32}
But, the LS method can not be applied to 
2D Ising universality transitions.

	\begin{figure}[t]
\includegraphics[width=\columnwidth]{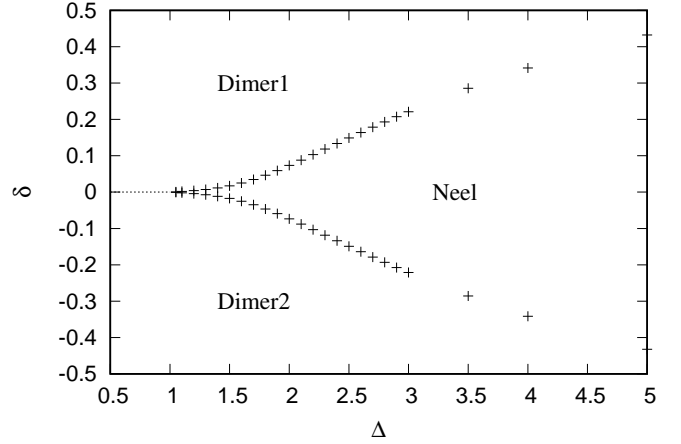}
\caption{Phase diagram in the 
	$\Delta-\delta$ plane.
	Dimer1-Dimer2 phase boundary is 
	the Gaussian universality and 
	Dimer-N${\rm \acute{e}}$el 
	phase boundaries are 
	the 2D Ising universality.
	We draw the 2D Ising universality transition lines 
	by using the L=24 numerical result 
	of yTBC-zTBC method, 
	noted by + and $\times$. 
	}
	\label{fig:pd}
\end{figure}
In this letter, 
as an example that has a multicritical point, 
we study an $S=\frac{1}{2}$ 
bond-alternating (BA) XXZ chain, 
	\begin{align}
	\hat{H}=
	\sum_{j}^{L}\left[1-(-1)^j\delta\right]
	\left(
	\hat{S}_j^x\hat{S}_{j+1}^x
	+\hat{S}_j^y\hat{S}_{j+1}^y
	+\Delta \hat{S}_j^z\hat{S}_{j+1}^z
	\right),
	\label{eq:xxz}
	\end{align}
where $L=2n$ ($n$ is integer). This model is 
	equivalent to the Ashkin-Teller model
	\cite{ashtel,2dis,srex}. 
This Hamiltonian with 
periodic boundary condition 
($\hat{{\bm S}}_{L+1}^{x}=
\hat{{\bm S}}_1^{x}$) 
is invariant under 
spin rotation around the z-axis 
($\hat{U}^z_{\theta}=
\exp{(i \theta \sum_j \hat{S}^z_{j})}$), 
spin reversal 
($\hat{U}^y_{\pi}=
\exp{(i \pi \sum_j \hat{S}^y_{j})}$) 
and two-sites translation 
($(\hat{T}_R)^2$ : 
$\hat{T}_R$ is one-site translation, 
$\hat{T}_R\hat{{\bm S}}_j 
\hat{T}_R^{-1}=
\hat{{\bm S}}_{j+1}$).
Corresponding eigenvalues are 
$M=\sum_j S^z_{j}$, 
$U^y_{\pi}=\pm 1$,
$q=4\pi n/L$ $(n=0,\cdots,L/2-1)$.
In addition, this model is symmetric 
about the sign of $\delta$, 
since the operation 
$\delta \rightarrow -\delta$ 
can be regarded as 
one-site translation 
$\hat{T}_R\hat{H}\hat{T}_R^{-1}$. 
At zero temperature, 
in the phase diagram of 
this model (\ref{eq:xxz}), 
the Gaussian universality transition line bifurcates 
into the two 2D Ising universality transition lines
at ($\Delta=1,\delta=0$) 
(Fig. \ref{fig:pd}).
In Dimer1 phase ($\delta>0$), 
the $2j-1,2j$ spins take a singlet pairing, 
and in Dimer2 phase ($\delta<0$), 
the $2j,2j+1$ spins take a singlet pairing, 
($j=1,2,\cdots,L$). 
In both phases, 
there exits a non-degenerate ground state 
with an energy gap. 
In N${\rm \acute{e}}$el phase, 
the ground states are doubly degenerate and 
the spin reversal symmetry is broken.
At the multicritical point ($\Delta=1,\delta=0$), called 
Ashkin-Teller multicritical 
point (AT point), 
a BKT transition occurs along $\delta=0$ 
where the correlation length 
diverges singularly. 
Near the AT point in $\Delta>1$,
 the two 2D Ising universality transition 
lines become
extremely close. 
Thus, 
since the finite size correction terms 
become very large,
one can not precisely calculate 
transition points near the AT point 
with the conventional methods. 
In our new method, 
getting an idea from
the Kramers-Wannier duality,\cite{duality} 
we use the two boundary conditions (BC's), 
which enable to calculate the 2D Ising 
universality transition points very accurately 
even near the AT point

	\section{Anisotropic Limit}
	We review that the BA XXZ model 
(\ref{eq:xxz}) 
 is identical to the TFI
model in the anisotropic limit,
\cite{xxz2}
then we shall discuss the boundary conditions. 
Firstly, we start in PBC.
We separate Eq.(\ref{eq:xxz}) 
to even bond and odd bond, 
\begin{align}
		\hat{H}=
		\beta
		\sum_{j}^{L/2}
		\left(
		\hat{S}_{2j}^x\hat{S}_{2j+1}^x
		+\hat{S}_{2j}^y\hat{S}_{2j+1}^y
		+\Delta\hat{S}_{2j}^z\hat{S}_{2j+1}^z
		\right)
		\nonumber \\
		+
		\sum_{j}^{L/2}
		\left(\hat{S}_{2j-1}^x\hat{S}_{2j}^x
		+\hat{S}_{2j-1}^y\hat{S}_{2j}^y
		+\Delta \hat{S}_{2j-1}^z
		\hat{S}_{2j}^z\right),
	\end{align}
$\beta=\frac{1-\delta}{1+\delta}$ and 
ignore the constant factor of the Hamiltonian.
In $\Delta \rightarrow \infty,
\beta \rightarrow 0,
\Delta\beta \sim O(1)$ limit, 
the 
$\Delta \hat{S}_{2j-1}^z\hat{S}_{2j}^z$ terms 
mostly contributes to the ground state energy. 
The ground state space is spanned by only 
\begin{eqnarray}
\begin{array}{l}
	\ket{\uparrow_{2j-1} \downarrow_{2j}}
	=\ket{\uparrow_j}',\\
	\ket{\downarrow_{2j-1} \uparrow_{2j}}
	=\ket{\downarrow_j}',
	\label{eq:sp1}
\end{array}
\end{eqnarray}
which are regarded as 
effective Ising spin states. 
The effective states and operators 
are denoted by $'$.
The first and second terms 
$
\hat{S}_{2j}^x\hat{S}_{2j+1}^x
+\hat{S}_{2j}^y\hat{S}_{2j+1}^y
$ 
vanish by $\beta\rightarrow0$.
The perturbative Hamiltonian is 
	\begin{align}
	\hat{H}_1=\sum_{j}^{L/2}
	\left(
	\beta\Delta\hat{S}_{2j}^z
	\hat{S}_{2j+1}^z\right)
	+
	\sum_{j}^{L/2}
	\left( \hat{S}_{2j-1}^x\hat{S}_{2j}^x
	+\hat{S}_{2j-1}^y\hat{S}_{2j}^y
	\right) \nonumber\\
	=
	\sum_{j}^{L/2}
	\left(
	\beta\Delta\hat{S}_{2j}^z
	\hat{S}_{2j+1}^z\right)
	+
		\frac{1}{2}
	\sum_{j}^{L/2}
	\left( 
		\hat{S}_{2j-1}^+\hat{S}_{2j}^-
	+\hat{S}_{2j-1}^-\hat{S}_{2j}^+
	\right).
	\label{eq:ptb1}
\end{align}
	We consider the first order 
	degenerate perturbation. 
	The first terms of (\ref{eq:ptb1}) are
	operating as 
\begin{align*}
\hat{S}_{2j+1}^z
	\ket{\uparrow_{j+1}}'
=
	\frac{1}{2}
	\ket{\uparrow_{j+1}}',\\
\hat{S}_{2j+1}^z
	\ket{\downarrow_{j+1}}'
=
	-\frac{1}{2}
	\ket{\downarrow_{j+1}}',\\
\hat{S}_{2j}^z
	\ket{\uparrow_j}'
=
	-\frac{1}{2}
\ket{\uparrow_j}',\\
\hat{S}_{2j}^z
\ket{\downarrow_j}'
=
	\frac{1}{2}
\ket{\downarrow_j}'.
\end{align*}
$\hat{S}_{2j}^z\hat{S}_{2j+1}^z$
can be regarded in the  effective space as 
$-\hat{S'}^{z}_{j}\hat{S'}^{z}_{j+1}$.
The second and third terms are operating as
\begin{align*}
\hat{S}_{2j-1}^+\hat{S}_{2j}^-
\ket{\downarrow_j}'
=
\ket{\uparrow_j}',\\
\hat{S}_{2j-1}^-\hat{S}_{2j}^+
\ket{\uparrow_j}'
=
\ket{\downarrow_j}'.
\end{align*}
Thus, 
	$
	\frac{1}{2}\left(
	\hat{S}_{2j-1}^+\hat{S}_{2j}^-
	+\hat{S}_{2j-1}^-\hat{S}_{2j}^+
	\right)$ 
can be regarded in the effective space as  
	$
	\frac{1}{2}
	\left(\hat{S'}_{j}^++\hat{S'}_{j}^-
	\right)=\hat{S'}_{j}^x
	$.
In summary, 
the effective Hamiltonian becomes 
\begin{align}
	\hat{H}'=\sum_j^{L/2}
		\left(
		-\beta\Delta
		\hat{S'}^z_j\hat{S'}^z_{j+1}
		+
		\hat{S'}^x_j
		\right).
\end{align}
By operating 
$\exp{(i \pi \sum_i^{L/2} \hat{S'}^z_j)}$, 
the effective Hamiltonian becomes 
the TFI model,
\begin{align}
	\hat{H}'=\beta\Delta\sum_j^{L/2}
		\left(
		-
		\hat{S'}^z_j\hat{S'}^z_{j+1}
		-\gamma
		\hat{S'}^x_j
		\right)
		\hspace{20pt}
		(\gamma\equiv\frac{1}{\beta\Delta}),
		\label{eq:ising}
\end{align}

	that has 
an order-disorder transition 
at $\gamma =1$, 
because of the Kramers-Wannier duality
\cite{duality}.
This Hamiltonian is invariant under 
 spin reversal and one-site translation.
We rewrite the Hamiltonian (\ref{eq:ising}), 
taking account of the boudary condition,
\begin{align}
	\hat{H}'=
		-\sum_j^{L/2-1}
		\hat{S'}^z_j\hat{S'}^z_{j+1}
		-
		g\hat{S'}^z_{L/2}\hat{S'}^z_{1}
		-
		\gamma
		\sum_j^{L/2}
		\hat{S'}^x_j
\end{align}
From exact solution\cite{apis},
for a finite system size $L$,
$E_0(L,g=1,U^y_{\pi}=-1)=
E_{0}(L,g=-1,U^y_{\pi}=1)+2(\gamma-1)$ 
is satisfied.
$E_0$ means a lowest state energy. 
At the transition point $\gamma=1$, 
the energies on the two BC's are crossing, 
\begin{align}
E_{0}(L,g=1,U^y_{\pi}=-1)=
E_{0}(L,g=-1,U^y_{\pi}=1).
	\label{eq:isap}
\end{align}
We can determinate the transition point of 
the TFI model by Eq. (\ref{eq:isap}).

	Next, we discuss what BC's of 
the BA XXZ model are corresponding
to the $g=1,-1$ of the TFI model.
For the BA XXZ model, 
we introduce z-axis twisted BC (zTBC),
\begin{align}
S_{L+1}^{x}=-S_1^{x},
S_{L+1}^{y}=-S_1^{y},
S_{L+1}^{z}=S_1^{z}. 
\label{eq:ztbc}
\end{align} 
The zTBC conserves 
the spin rotational symmetry and 
the spin reversal symmetry, 
but 
breaks the two-sites translational symmetry. 
In the anisotropic limit, 
the effective Hamiltonian becomes 
the $g=1$ TFI model, 
since the x,y-direction 
boundary terms vanish.

And, we introduce y-axis twisted BC (yTBC), 
\begin{align}
S_{L+1}^{x}=-S_1^{x},
S_{L+1}^{y}=S_1^{y},
S_{L+1}^{z}=-S_1^{z}.
\label{eq:ytbc}
\end{align}
The yTBC (\ref{eq:ytbc}) 
conserves the spin reversal symmetry, 
but breaks 
the two-sites translational symmetry 
and the spin rotational symmetry.
About the last point, 
since the boundary terms become 
\begin{align*}
	-\hat{S}^x_{L} \hat{S}^x_{1} 
	+
	\hat{S}^y_{L} \hat{S}^y_{1} 
	-\Delta
	\hat{S}^z_{L} \hat{S}^z_{1}
	=
	-\frac{1}{2}\left(
	\hat{S}^+_{L} \hat{S}^+_{1} 
	+
	\hat{S}^-_{L} \hat{S}^-_{1} 
	\right)
	-\Delta
	\hat{S}^z_{L} \hat{S}^z_{1},
\end{align*}
thus $M$ can not be conserved. 
But, 
the Hamiltonian is particularly invariant  
under $\pi$-rotation around z-axis
$\hat{U}^z_{\pi}=\exp{(i \pi \sum_j \hat{S}^z_{j})}
=
(-1)^{\hat{M}}
$. 
So, a parity of a total magnetization 
${P}_M=(-1)^{M}$ 
is a conserved quantity.
In the anisotropic limit, 
the z-direction boundary terms remain minus, 
$-\Delta \hat{S}_L^z\hat{S}_{1}^z$.
So, 
the effective Hamiltonian becomes 
the $g=-1$ TFI model.

	Consequently, 
PBC and zTBC of the BA XXZ model 
correspond to $g=1$ of the TFI model,
and the yTBC corresponds to $g=-1$.
Thus, Eq. (\ref{eq:isap}) of the TFI model 
is extended to the BA XXZ model, 
\begin{align}
E_{0}^{PBC}(M=0,U^y_{\pi}=-1)=
E_{0}^{yTBC}(M=even,U^y_{\pi}=1),
\label{eq:py}
\end{align}
(hereafter we call the yTBC-PBC method) 
or
\begin{align}
E_{0}^{zTBC}(M=0,U^y_{\pi}=-1)=
E_{0}^{yTBC}(M=even,U^y_{\pi}=1),
\label{eq:zy}
\end{align}
(we call the yTBC-zTBC method).
The above quantum numbers are for $\delta>0$. 
In Table \ref{tab:qnm}, 
the quantum number are summarized for 
$\delta>0$ and $\delta<0$. 
Although the finite size corrections vanish 
in the anisotropic limit, 
they remain in the finite $\Delta$ region 
because of a perturbation in the process 
from Eq. (2) to Eq. (7). 
However, we consider that 
the relation of Eq. (11) and (12) are supported
by the 2D Ising universality class or 
the c=1/2 conformal field theory. 
We shall discuss on this point in a future paper.
\begin{table}[t]
	\caption{Quantum numbers of 
	the eigenstates 
	of PBC, zTBC and yTBC for $\delta>0$. 
	The case for $\delta<0$ is denoted by (). 
	The states used for  
	yTBC-zTBC method
	are denoted with $\bullet$ 
	and yTBC-PBC method with $\circ$.
	}
\begin{tabular}{|l|c|c|}
\hline
& $M$ & $U^y_{\pi}$ \\ \hline
$\textcolor{white}{\bullet} \circ$
PBC & 0 & -1 (-1)\\\hline
$\bullet \textcolor{white}{\circ}$
zTBC & 0 & -1 (1)\\\hline
$\bullet \circ$
yTBC & even (odd) & 1 (1)\\
\hline
\end{tabular}
	\label{tab:qnm}
\end{table}

	\section{Isotropic Limit}
	\begin{figure}[t]
\includegraphics[width=\columnwidth]{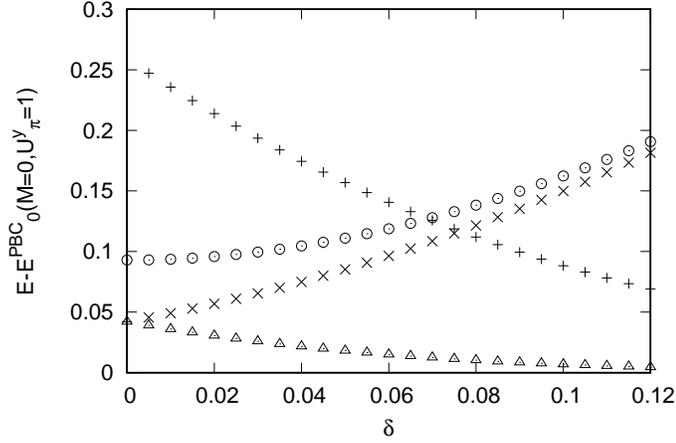}
\caption{
The energies of each BC  
for L=14. 
The value of $\Delta$ is fixed at $2.0$ 
and $\delta$ is changed. 
$\bigcirc$ is $E^{PBC}_0(M=0,U^y_{\pi}=-1)$,
$\times$ is $E^{zTBC}_0(M=0,U^y_{\pi}=-1)$, 
$\triangle$ is $E^{zTBC}_0(M=0,U^y_{\pi}=1)$, 
$+$ is $E^{yTBC}_0(M=even,U^y_{\pi}=1)$. 
The PBC lowest energy 
$E^{PBC}_0(M=0,U^y_{\pi}=1)$ is subtracted from 
each energy.
}
\label{fig:cross}
\end{figure}
On the self dual line ($\delta=0$), 
the Hamiltonian with zTBC becomes, 
\begin{align}
\hat{H}=
\sum_{j}^{L-1}
\left(
\hat{S}_j^x\hat{S}_{j+1}^x
+\hat{S}_j^y\hat{S}_{j+1}^y
+\Delta \hat{S}_j^z\hat{S}_{j+1}^z\right)\nonumber\\
-\hat{S}_L^x\hat{S}_{1}^x
-\hat{S}_L^y\hat{S}_{1}^y
+\Delta \hat{S}_L^z\hat{S}_{1}^z.
\label{eq:xxzztbc}
\end{align}
We define 
$\pi$ spin rotation at $j$ site about z-axis 
as 
$\hat{u}^z_j=\exp{(i \pi \hat{S}^z_j)}$. 
Since the Hamiltonian (\ref{eq:xxzztbc}) is invariant 
under $\hat{T}_R\hat{u}_L^z$, 
\begin{align}
\hat{T}_R\hat{u}^z_L
\hat{H}
\ket{U^y_{\pi}}
=\hat{H}
\hat{T}_R\hat{u}^z_L\ket{U^y_{\pi}}.
\end{align}
The commutation relation between 
$\hat{U}^y_{\pi}$ and $\hat{u}^z_L$ is
\begin{align}
\hat{U}^y_{\pi}\hat{u}^z_L=&
\exp{(i \pi \sum_j \hat{S}^y_j)}
\exp{(i \pi \hat{S}^z_L)}\nonumber\\
=&
\exp{(-i \pi \hat{S}^z_L)}
\exp{(i \pi \sum_j \hat{S}^y_j)}\nonumber\\
=&
\exp{(-2 i \pi \hat{S}^z_L)}
\hat{u}^z_L\hat{U}^y_{\pi}.
\end{align}
When S is a half-integer, 
the eigenvalue of $\hat{S}_L^z$ is a half-integer, 
\begin{align}
\hat{U}^y_{\pi}\hat{u}^z_L=
-\hat{u}^z_L\hat{U}^y_{\pi}.
\end{align}
So, 
\begin{align}
\hat{U}^y_{\pi}\hat{T}_R\hat{u}_L^z
\ket{U^y_{\pi}=1}
=&-\hat{T}_R\hat{u}_L^z\hat{U}^y_{\pi}
\ket{U^y_{\pi}=1}\nonumber\\
=&-\hat{T}_R\hat{u}_L^z\ket{U^y_{\pi}=1}\\
\equiv&-\ket{U^y_{\pi}=-1}.\nonumber
\end{align}
Consequently, 
$\ket{U^y_{\pi}=1}$ and $\ket{U^y_{\pi}=-1}$ 
are degenerate for an arbitrary $L$, 
	\begin{align}
		E_0^{zTBC}(L,M=0,U^y_{\pi}=-1)=
		E_0^{zTBC}(L,M=0,U^y_{\pi}=1), 
	\label{eq:del1yz}
\end{align}
as you can see in Fig.\ref{fig:cross} 
at $\delta=0$.

      Furthermore, 
on the isotropic point ($\Delta=1$), 
	the yTBC is equivalent to zTBC. 
	Thus, replacing zTBC of 
	the right side of 
	Eq. (\ref{eq:del1yz}) with yTBC, 
	\begin{align*}
		E_0^{zTBC}(L,M=0,U^y_{\pi}=-1)=
		E_0^{yTBC}(L,M=even,U^y_{\pi}=1).
	\end{align*}
	Consequently, 
	since the correction terms 
	vanish at the AT point, 
	the yTBC-zTBC method (\ref{eq:zy})
	can suppress the effect of the AT point. 
	In contrast, 
	the correction terms of 
	the yTBC-PBC method 
	remains	at the AT point.

	\section{Numerical Calculation}
       We use the exact diagonalization method
to calculate the energies of each BC.
We determine 
the energy crossing point $\delta_c$, 
by fixing $\Delta$ and changing $\delta$,  
 on L=10,$\cdots,24$, 
as Fig.\ref{fig:cross}. 
Note that the PBC ground state and 
the zTBC lowest state are not 
related with the Kramers-Wannier duality (\ref{eq:isap}).
We show the phase diagram 
Fig.\ref{fig:pd}. 
\begin{figure}
	\includegraphics[width=\columnwidth]{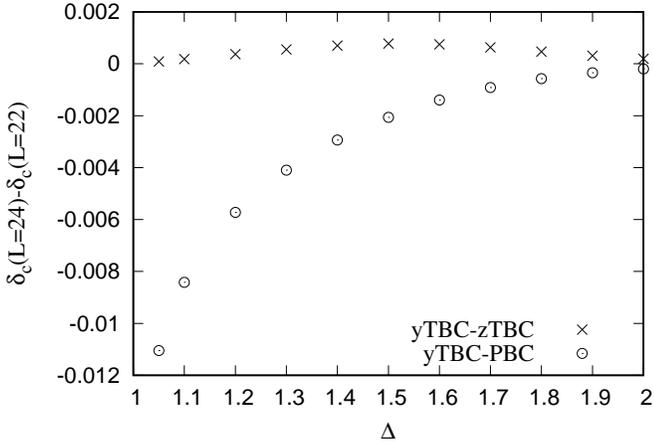}\\
	\caption{
The size difference of the crossing point 
$\delta_c$ L=24 and L=22. 
$\bigcirc$ is
the yTBC-PBC method, 
$\times$ is 
the yTBC-zTBC method.
	}
	\label{fig:vld}
\end{figure}
\begin{figure}
\includegraphics[width=\columnwidth]{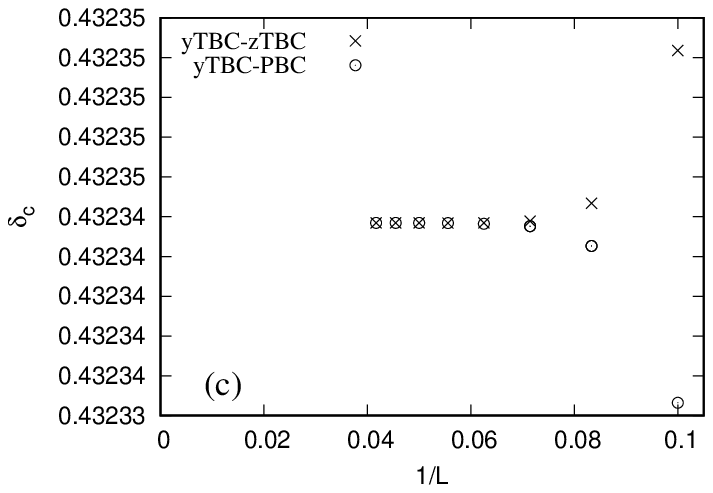}\\
\includegraphics[width=\columnwidth]{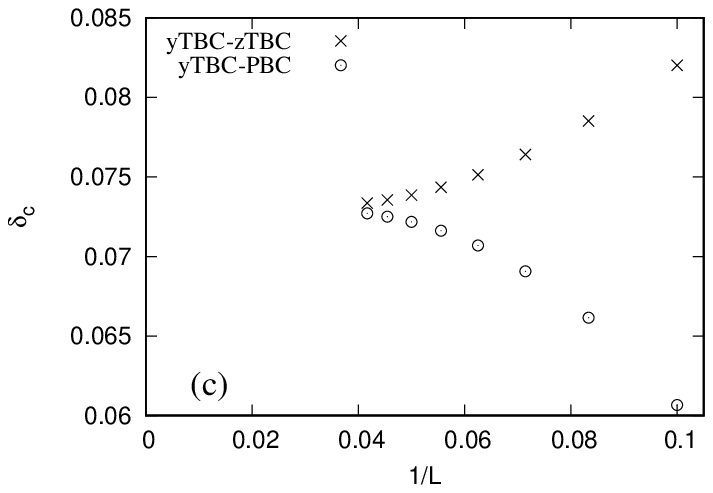}\\
\includegraphics[width=\columnwidth]{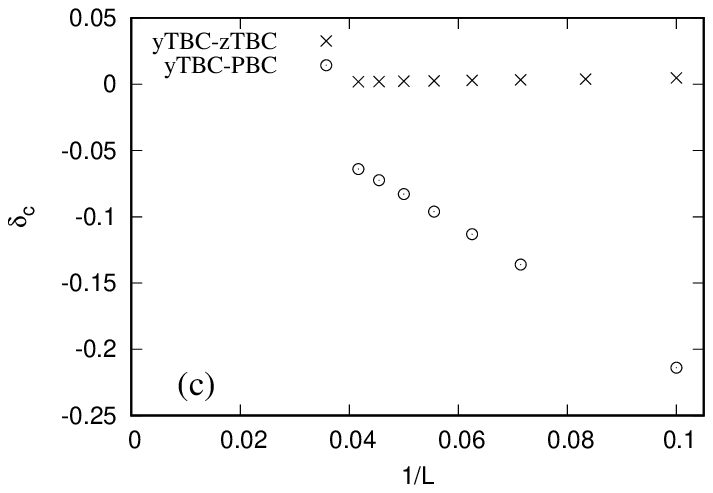}\\
	\caption{
The size dependence of 
the crossing point 
$\delta_c$ 
of 
the yTBC-PBC method ($\bigcirc$) and 
the yTBC-zTBC method ($\times$), 
	(a) in $\Delta=5.0$,
	(b) in $\Delta=2.0$,
	(c) in $\Delta=1.1$. 
	}
	\label{fig:yzyp}
\end{figure}

	The size dependence shows in 
Fig. \ref{fig:yzyp}. 
The crossing points $\delta_c(L)$ 
by the yTBC-zTBC method
are monotonically decreasing with size L, 
whereas those of the yTBC-PBC method are 
monotonically increasing. 
As Fig. \ref{fig:yzyp}(a), 
in the large $\Delta$ region, 
the yTBC-zTBC method and the yTBC-PBC method 
are rapidly converged. 
Next, near the AT point, 
as Fig. \ref{fig:yzyp}(c),
the finite size corrections of the yTBC-PBC method 
become very large.  
In contrast, the yTBC-zTBC method shows 
a well convergence.

To compare the two methods 
from another viewpoint, 
we show the size difference of 
the crossing point  
$\delta_c(L=24)-\delta_c(L=22)$ 
in Fig. \ref{fig:vld}. 
For large $\Delta$, 
the two methods are almost the same and 
the size difference vanish. 
Near the AT point,  
the size difference of 
the yTBC-PBC method becomes very large, 
but 
the yTBC-zTBC method approaching to zero.

        \section{Conclusion}
        Using the yTBC-PBC method (\ref{eq:py}) or 
the yTBC-zTBC method (\ref{eq:zy}),
we can numerically calculate 
2D Ising universality transition points. 
We actually calculate 
the transition lines 
of S=1/2 BA XXZ model. 
As expected, 
the yTBC-zTBC method reduces 
the finite size effects near 
the multicritical point, 
since the finite size correction terms 
vanish at AT point. 
About critical exponents and 
the universality class, 
we shall describe them in a future paper. 
Furthermore, we verify the accuracy 
by comparing the yTBC-zTBC numerical result 
with the result of 
renormalization group theory\cite{rg}. 
We expect our method can be applied 
to several quantum spin models.

	\section{Acknowledgement}
	Our calculation program used 
in the exact diagonalization
is TITPACK Ver.2 
coded by H. Nishimori \cite{tit}. 
A modification of the calculation program 
for yTBC is assisted by T.Mukai.

\end{document}